
\magnification 1200
\font\abs=cmr9

\def\uno{{\bf 1}}

\def\fraz#1#2{{\strut\displaystyle #1\over\displaystyle #2}}
\def\dual#1#2{\langle #1 , #2\rangle}

\def\esp#1{e^{\displaystyle#1}}
\def\ii#1{\item{$\phantom{1}[#1]~$}}
\def\jj#1{\item{$[#1]~$}}

\def\ee{\varepsilon}
\def\fun#1{{\sl Fun}\left(#1\right)}
\def\funq#1{{\sl Fun}_q\left(#1\right)}
\hsize= 15 truecm
\vsize= 22 truecm
\hoffset= 1. truecm
\voffset= 0.3 truecm

\baselineskip= 14 pt
\footline={\hss\tenrm\folio\hss} \pageno=1

\centerline{\bf EXPONENTIAL  MAPPING FOR NON SEMISIMPLE  QUANTUM GROUPS.}

\bigskip\bigskip
\centerline{
{\it
F.Bonechi, E.Celeghini, R.Giachetti ${}^1$, C.M.Pere\~na
\footnote{${}^{\dag}$}{On leave from Departamento de
F\'\i sica Te\'orica, Universidad Complutense, 28040 Madrid, Spain},
E.Sorace and M.Tarlini.}}
\bigskip
{}~Dipartimento di Fisica, Universit\`a di Firenze and INFN--Firenze,
\footnote{}{\smallskip\hskip -.85truecm E--mail:
TARLINI@FI.INFN.IT
\hfill DFF 192/9/93 Firenze}

${}^1$Dipartimento di Matematica, Universit\`a di Bologna and INFN--Firenze.
\bigskip
\bigskip

{\bf Abstract.} {\abs The concept of universal $T$ matrix, recently
introduced by Fronsdal and Galindo [1] in the framework of quantum
groups, is here
discussed as a generalization of the exponential mapping.
New examples
related to inhomogeneous quantum groups of physical interest are developed,
the duality calculations are explicitly presented and it is found
that in some cases the universal $T$ matrix, like for Lie groups, is
expressed in terms of usual exponential series.}

\medskip
\noindent PACS 02.20.+b; 03.65.Fd \hfill\break

\bigskip
\bigskip

\noindent {\bf 1. Introduction.}

\medskip

The relations of the quantum algebra $\funq{SU(2)}$ generated by the
elements $a$, $b$, $c$ and $d$ have the remarkable property of being
preserved under matrix multiplication [2]. This means that if we
define
$$T'=\pmatrix{a'&b'\cr c'&d'\cr}\ \ \ \ \ \ \ \ \
T''=\pmatrix{a''&b''\cr c''&d''\cr}\ ,
$$
where $(a',\ b',\ c',\ d'\ )$ and $(a'',\ b'',\ c'',\ d''\ )$ are two
mutually commuting sets of elements satisfying the relations of
$\funq{SU(2)}$, then the variables $(a,\ b,\ c,\ d\ )$ defined by
$$ T=\pmatrix{a&b\cr c&d\cr}\,\equiv\, T'\;T''
$$
also satisfy the same relations.
This property can be formalized by defining a comultiplication $\Delta$
according to
$$\Delta(T)=T\;{\dot \otimes}\;T,\ \eqno{(1)}$$
where ${\dot \otimes}$ denotes matrix multiplication and tensor product
of the $C^*$-algebras of the noncommutative representative functions,
[2]. For $q=1$ the matrix $\Delta(T)$, with elements
in $\otimes^2\fun{SU(2)}$, gives rise to the ordinary group composition.
The inverse matrix then defines a second operation on the
elements $(a,\ b,\ c,\ d\ )$, namely the antipode
$$S(T)=T^{-1}\ .\eqno{(2)}$$

The antipode and comultiplication together with counit, unit and
multiplication are collected into the Hopf algebra $\funq{SU(2)}$.

These properties can be abstracted and generalized without  referring
any more to matrices. To show this, we
consider the quantization $U_q(g)$ of the universal
enveloping algebra of a Lie algebra $g$ with product,
coproduct and antipode defined by
$$ m(X_A\otimes X_B)\,=\,f_{AB}^C\; X_C\ ,$$
$$\Delta(X_A)\,=\,h_A^{BC}\; X_B\otimes X_C\ ,$$
$$S(X_A)\,=\,s_A^B\;X_B\ ,$$
where the sum over repeated indices is assumed and
where $X_A$ are the elements of a basis of $U_q(g)$, as, for instance, the
Poincar\'e--Birkhoff--Witt basis.
If $g$ is a compact form of a semisimple Lie algebra, it is well known that
the Tannaka theory establishes a duality between the universal enveloping
algebra and the Hopf algebra of the representative functions. This kind of
duality has been studied also at the quantum level [3] so that we can speak
of a compact quantum group $\funq{G}$ satisfying the relations
$$ m^*(x^C)\,=\,f_{AB}^C\; x^A\otimes x^B\ ,$$
$$\Delta^*(x^B\otimes x^C)\,=\,h_A^{BC}\; x^A\ ,$$
$$S^*(x^B)=(s^{-1})^B_A\;x^A\ ,$$
where $\{x^A\}$ is a basis of $\funq{G}$ such that
$\dual{x^A}{X_B}=\delta^A_B$.
For non compact Lie algebras the duality is more delicate and the
functions vanishing at infinity must be determined [4].

In this scheme, independent of the representation, the object that takes
the place of the matrix $T$, called the universal $T$-matrix
and denoted by the same letter, is given by summing the
tensor products of all the corresponding elements of a pair of dual
bases [1]:
$$T=x^A\otimes X_A\,.$$
The structure of $T$ and the Hopf algebra operations naturally suggest
two kinds of mappings, the first one using the multiplication of $U_q(g)$,
$$T\;{\dot \otimes}\;T=(x^A\otimes x^B)\otimes m(X_A\otimes X_B)\ ,
\eqno{(3)}$$ the other being obtained from the multiplication of $\funq{G}$,
$$T\;{\ddot \otimes}\;T=\Delta^*(x^A\otimes x^B)\otimes (X_A\otimes X_B)\
.\eqno{(4)}$$
It is straightforward to see that the duality relations yield the equalities
$$m^*(x^A)\otimes X_A=T\;{\dot \otimes}\;T\ ,\quad\quad
x^A\otimes \Delta(X_A)=T\;{\ddot \otimes}\;T\ ,\eqno{(5)}$$
and
$$S^*(x^A)\otimes X_A\; =\; x^A\otimes S(X_A)\; =\; T^{-1}\ ,
\eqno{(6)} $$
where $T^{-1}$ is defined so to have
$$m^*\,(T{\dot \otimes} T^{-1})=
 m^*\,(T^{-1}{\dot \otimes} T)=\uno~~~{\rm and}~~~
 m\,(T{\ddot \otimes} T^{-1})=
 m\,(T^{-1}{\ddot \otimes} T)=\uno.$$

For the sake of clarity let us consider the explicit example of a compact
Lie algebra $g$ with corresponding Lie group $G$
and representative functions  $\fun{G}$.
If $X_k\ $, $(k=1,\dots n)$, are the generators of the Lie algebra, a
basis of the universal enveloping algebra is of the form
$X_A=X_1^{a_1}X_2^{a_2}\cdots X_n^{a_n}$.
The dual elements $x^A\in\fun{G}$ are then
$x^A=x_1^{a_1}x_2^{a_2}\cdots x_n^{a_n}/(a_1!a_2!\cdots a_n!)\ $
where $\dual{x_k}{X_\ell}=\delta_{k \ell}$. Therefore the universal $T$
matrix results in
$$T=\sum_{a_1}{\fraz{x_1^{a_1}\otimes X_1^{a_1}}{a_1!}}\cdots
    \sum_{a_n}{\fraz{x_n^{a_n}\otimes X_n^{a_n}}{a_n!}}\ =\
            \esp{\;x_1\otimes X_1}\;\cdots\; \esp{\;x_n\otimes X_n}\ .$$

It appears that the evaluation of $T$ on an element of the group $G$
reproduces that element expressed by means of the exponential mapping
between $g$ and $G$ and therefore in the case
of Lie groups the universal $T$ matrix can be regarded as a resolution
of the identity mapping of $G$ into itself. This point of view must be
slightly modified in quantization: recalling that the evaluation
determines a character on the algebra of the representative functions, we
see that this character reproduces itself when applied to the universal
$T$ matrix, despite the
fact that $x^A$ are now elements of a noncommutative $C^*$-algebra.

In the standard framework [2] the quantum relations are obtained using the
matrices $T\otimes \uno$ and $\uno\otimes T$. Accordingly we define
$$T_1=x^A\otimes(X_A\otimes \uno)\ ,~~~~~~~~~~
T_2=x^B\otimes(\uno\otimes X_B)\ ,$$
so that the products $T_1T_2$ and $T_2T_1$ read
$$T_1T_2=\Delta^*(x^A\otimes x^B)\otimes (X_A\otimes X_B)\ ,$$
$$T_2T_1=\Delta^*(x^B\otimes x^A)\otimes (X_A\otimes X_B)\ ,$$
and, as shown before, they can be expressed in the form
$$T_1T_2=x^C\otimes \Delta(X_C)\ , \quad~~~\quad T_2T_1=x^C\otimes \sigma
\Delta(X_C)\ ,$$
where $\sigma( X\otimes Y)=Y\otimes X$. We then see that when an $R$-matrix
does exist, its defining property $R\Delta R^{-1} = \sigma \Delta$
gives immediately the algebraic relation
$$R\ T_1\ T_2\,=\,T_2\ T_1\ R\,,\eqno(7)$$
which, when represented, reproduces the well known quantization prescription,
[2].

In the next section we briefly summarize the results for the
quantum group $SU_q(2)$, which, up to minor additions on the antipode,
are contained in [1]. The purpose for so doing is twofold: first we find
it useful to give a developed example of the way in which the universal
$T$ matrix works; secondly we want to establish explicit relations that
will be relevant to discuss the universal $T$ matrix for some inhomogeneous
quantum groups that are related to $SU_q(2)$
and useful for physical applications, namely
$H_q(1)$ [5], $E_q(2)$ [6] and $\Gamma_q(1)$ [7].
These will be presented in subsequent sections where we shall see that
the  $T$ operator, expressed in terms of $q$--exponentials for $SU_q(2)$,
in some cases and in an appropriate basis is simply given
by a product of exponentials.

\bigskip\bigskip

\noindent {\bf 2. The universal $T$ matrix for $SU_q(2)$.}

\medskip
Starting from the usual generators $J_+$, $J_-$ and $J_3$ of $SU_q(2)$
we define
$$E=\esp{z\,J_3/2}\ J_+\ ,~~~~~\quad F=\esp{-z\,J_3/2}\ J_-\ ,$$
that satisfy the commutation relations
$$\eqalign{
{}&[J_3,E]\;=\;E\ ,\quad\quad [J_3,F]\;=\;-F\ ,\cr
      {}&[E,F]\;=\fraz{2\; \sinh{(z J_3)}}{1-\esp{-z}}\ .\cr}$$
The coproduct and antipodes have now the form
$$\eqalign{\Delta E\;=\; \uno\otimes E + E &\otimes \esp{z J_3}\ ,~~~~~~~~
           \Delta F\;=\; \esp{-z J_3}\otimes F + F \otimes \uno\ ,\cr
         {}&\Delta J_3\;=\; \uno\otimes J_3 + J_3 \otimes \uno\ \cr
S(E)\,=\,-e^{-z J_3}\;E\ , &\quad\quad S(F)=\,-e^{z J_3}\;F\ ,\quad
\quad S(J_3)\,=\, -J_3\ .\cr}\eqno{(8)}$$

In order to find the $T$ operator we must determine the dual $\funq{SU(2)}$.
We thus begin by defining the elements $\,\phi,\,\gamma,\,\eta\,$ dual
to the generators $\,E,\,F,\,J_3\,$ satisfying
$$\dual{\phi}{F}=1\ ,\quad\quad \dual{\gamma}{J_3}=1\ ,\quad\quad
\dual{\eta}{E}=1\ .\eqno{(9)}$$
The three elements $\,\phi,\,\gamma,\,\eta\,$ generate  $\funq{SU(2)}$ as
an algebra [1] and satisfy
$$[\phi,\eta]=0\ , \quad\quad [\gamma,\phi]=-z\,\phi\ ,\quad\quad
[\gamma,\eta]=-z\, \eta\ .\eqno{(10)}$$
The coproducts and antipodes are
$$\eqalign{
m^*(\phi)=&\;\phi\otimes \uno + (e^{-\gamma/2}\otimes \phi)
(\uno\otimes \uno + \eta\otimes \phi)^{-1}
(e^{-\gamma/2}\otimes \uno)\ ,\cr
m^*(\eta)=&\;\uno\otimes \eta + (\uno\otimes e^{-\gamma/2})
(\uno\otimes \uno + \eta\otimes \phi)^{-1}
(\eta\otimes e^{-\gamma/2})\ ,\cr
m^*(\gamma)=&\;\uno\otimes \gamma + \gamma\otimes \uno -
2\; z\; \sum_n{\fraz{(-\eta\otimes \phi)^n}{1-e^{-z\, n}}}\ .\cr
}\eqno{(11)}
$$
$$\eqalign{{}&S^*(\phi)=-(e^{-\gamma/2}+\eta e^{\gamma/2} \phi)^{-1}
e^{\gamma/2} \phi\ , \cr
{}&S^*(\eta)=-\eta e^{\gamma/2} (e^{-\gamma/2}+\eta e^{\gamma/2}
\phi)^{-1}\ , \cr
{}&S^*(e^{\gamma/2})=\; e^{-\gamma/2}+\eta e^{\gamma/2} \phi\ .
\cr}\eqno{(12)}$$

The elements
$$X_{jk\ell}\,=\,F^j\;J_3^k\;E^\ell$$
define a basis of $U_q(su(2))$. A direct verification shows that the
corresponding dual basis in  $\funq{SU(2)}$ is given by
$$x^{jk\ell}\,=\,\fraz{\phi^j}{[j]_z!}\ \fraz{\gamma^k}{k!}\
\fraz{\eta^\ell}{[\ell]_{-z}!}\ ,$$
where $[n]_z=(e^{z\, n}-1)/(e^z-1)$ and
$[n]_z!=[n]_z\,[n-1]_z\,\dots [2]_z\,[1]_z$.

The universal $T$ matrix is then [1]
$$T\;=\;e^{\;\phi\otimes F}_z\ \ e^{\;\gamma\otimes J_3}\ \ e^{\;\eta\otimes
E}_{-z}\ ,
\eqno{(13)}$$
where $\esp{A}_z=\sum_i{\fraz{A^i}{[i]_z!}}\ $ (to compare with [1]
$\ q=q'=e^{z/2}$).

The content of equations (5) can now be made explicit. Indeed, if for
any element $\xi\in\funq{SU(2)}$ we resume the initial notations
$\xi'=\xi\otimes\uno$, $\xi''=\uno\otimes\xi$ an we rewrite equations (11)
accordingly, we get the suggestive result

$$\eqalign{
{}&\Bigl(e^{ \; \phi'\otimes F }_z\ \
e^{ \; \gamma'\otimes J_3 }\ \
e^{ \; \eta'\otimes E }_{ -z }\Bigr)\ \
\Bigl(e^{ \; \phi''\otimes F }_z\ \
e^{ \; \gamma''\otimes J_3 }\ \
e^{ \; \eta''\otimes E }_{ -z}\Bigr)\ =\cr
{}&e^{ \; m^*(\phi)\otimes F }_z\ \
e^{ \; m^*(\gamma)\otimes J_3 }\ \
e^{ \; m^*(\eta)\otimes E }_{ -z }\ =\cr
{}&e^{ \; (\phi' + e^{-\gamma'/2}\phi''
(1 + \eta' \phi'')^{-1} e^{-\gamma'/2})
\otimes F }_z\ \
e^{ \; (\gamma' + \gamma'' -
2z\,\sum_n (-\,\eta' \phi'')^n\,/\,(1-e^{-zn}))
\otimes J_3 }\ \cdot\cr
{}&\phantom{e^{ \; (\phi' + e^{-\gamma'/2}\phi''
(1 + \eta' \phi'')^{-1} e^{-\gamma'/2})
\otimes F }_z}~~~
e^{ \; (\eta'' +  e^{-\gamma''/2} (1 + \eta' \phi'')^{-1}
\eta' e^{-\gamma''/2})
\otimes E }_{ -z }\ .\cr
}\eqno(14)$$

Obviously, in the limit $z\rightarrow 0$ we recover the exponential mapping
and the Lie group multiplication. Moreover equation (14) gives a very neat
example of the conditions posed by the Friedrichs theorem [8]: indeed the
noncommutativity of $\funq{SU(2)}$
and the presence of non standard exponentials
is needed to compensate the fact that the generators of the quantum
algebra are no more primitive. Additional peculiarities
are also connected with the antipode, but we shall present
them for the Heisenberg quantum group with the explicit calculations, which,
in that case, are much simpler. We finally observe that
the expression for $m^*(\gamma)$ depends on $z$ and that the limit
$z\rightarrow 0$ gives the classical composition law
$$\lim_{z\to 0}\ m^*(\gamma)\ =\ \uno\otimes \gamma + \gamma\otimes
\uno + 2 \log(\uno\otimes \uno + \eta\otimes\phi)\ .$$

\bigskip
\bigskip

\noindent {\bf 3. The exponential mapping for $H_q(1)$.}

\medskip

In reference [5] a quantum deformation of the Heisenberg group has been
determined. With a slight change in the definitions with respect to [5]
the commutation relations of the generators $a_-$, $a_+$ and $H$
of the quantum Heisenberg algebra can be written in the form
$$[a_-,a_+]=\fraz{\sinh(w\;H)}{w}\ ,\quad\quad [H,\;\cdot\;]=0\ .$$
The corresponding coproducts read
$$\eqalign{\Delta(a_-)=\;&\uno\otimes a_- + a_-\otimes \esp{w\;H}\ ,\cr
  \Delta(a_+)=\;&\esp{-w\;H}\otimes a_+ + a_+\otimes \uno \ ,\cr}$$
$H$ being primitive, while the antipodes are
$$S(a_-)=-e^{w H}\;a_-\,,~~~~~~~ \ S(a_+)=-e^{-w H}\;a_+\,,~~~~~~~
\ S(H)=-H\,.$$

In order to determine the dual structure $\funq{H(1)}$ we consider the
generators $\ \alpha\ ,\ \beta\ ,\ \delta\ $ satisfying
$\dual{\alpha}{a_-}=1\ ,\,\dual{\beta}{H}=1\ ,\,\dual{\delta}{a_+}=1$
with commutation relations, coproducts and antipodes given by:
$$[\alpha,\delta]=0\ ,\quad\quad [\beta,\alpha]=-w\;\alpha\ ,\quad\quad
[\beta,\delta]=-w\;\delta\ ,$$
$$\eqalign{
  m^*(\alpha)=&\;\alpha\otimes \uno + \uno\otimes \alpha\ ,\cr
  m^*(\beta)=&\;\beta\otimes \uno + \uno\otimes \beta +
  \alpha\otimes\delta\ ,\cr
  m^*(\delta)=&\;\delta\otimes \uno + \uno\otimes \delta\ ,\cr}
\eqno(15)$$
$$S^*(\alpha)=-\alpha\ , \quad\quad
S^*(\beta)=-\beta+\alpha\delta\ ,\quad\quad
S^*(\delta)=-\delta\ .$$
A direct calculation shows that
$\dual{\delta^a\;\beta^b\;\alpha^c}{a_+^d\;H^e\;a_-^f}=
a!\,\delta_{ad}\;b!\,\delta_{be}\;c!\,\delta_{cf}$,
so that the $T$ matrix results in:
$$
T=e^{\;\delta\otimes a_+}\ \
e^{\;\beta\otimes H}\ \
e^{\;\alpha\otimes a_-}\ .$$

Unlike the case of $SU_q(2)$ and similarly to what occurs for
Lie algebras, the universal $T$ matrix is now expressed in terms of simple
exponentials: however, as already observed at the end of the previous
section, the non primitive nature  of the quantum generators
$a_-$, $a_+$ and $H$ must be compensated by
the non vanishing commutators of the $C^*$--algebra elements $\delta$,
$\alpha$ and $\beta$ in order to reproduce the same `` group composition''
of the coordinates as given in (15). Let us also show directly
that the inverse of the universal $T$ matrix
cannot be expressed in terms of exponentials of the
antipodes: indeed, from (6),
$$\eqalign{T^{-1}&=S^*(\fraz{\delta^a}{a!}\;\fraz{\beta^b}{b!}\;
\fraz{\alpha^c}{c!})\otimes a_+^a\;H^b\;a_-^c\cr
{}&=\fraz{1}{c!}S^*(\alpha)^c\;\fraz{1}{b!}S^*(\beta)^b\;
\fraz{1}{a!}S^*(\delta)^a\otimes a_+^a\;H^b\;a_-^c\cr
{}&=\fraz{\delta^a}{a!}\;\fraz{\beta^b}{b!}\;\fraz{\alpha^c}{c!}\otimes
S(a_+^a\;H^b\;a_-^c)\cr
{}&=\fraz{\delta^a}{a!}\;\fraz{\beta^b}{b!}\;\fraz{\alpha^c}{c!}\otimes
S(a_-)^c\;S(H)^b\;S(a_+)^a\ .\cr}
$$
If we reorder the terms in this expressions so to reconstruct exponential
series, we find
$$T^{-1}=e^{-\alpha\otimes a_-}\ \ e^{-\beta\otimes H}\ \ e^{-\delta\otimes
a_+}\ ,
$$
namely the obvious expression that, however, has not the chosen ordering
of the three exponential factors: this, indeed,
is related to the deep question of what should be
taken as a quantum analogue of the classical Baker-Campbell-Hausdorff
formula [10].

It was shown in [5] that the results for $H_q(1)$ can be obtained by
contracting the quantum algebra $SU_q(2)$. It is interesting to observe
that this procedure holds also at the level of the quantum
group $\funq{SU(2)}$ and therefore the $T$ matrix itself can be obtained
by contraction. Indeed the rescaling $a_-=\ee^{1/2}\;E$,
$a_+=\ee^{1/2}\;F$,
$H=\ee\; 2 J_3$, with $w=\ee^{-1} z/2$ reproduces in the limit
$\ee\rightarrow 0$ the quantum algebra $H_q(1)$. In order to maintain
the pairing relations (9) we have to define $\alpha=\ee^{-1/2}\eta$,
$\beta=\ee^{-1}\gamma/2$ and $\delta=\ee^{-1/2}\phi$: the relations of
$\funq{H(1)}$ are simply obtained by using this rescaling on $\funq{SU(2)}$
and taking $\ee \rightarrow 0$.
The $T$ matrix for $H_q(1)$ is then calculated from that of $SU_q(2)$
by taking the limit $\ee\rightarrow 0$: because of this limit it is clear
that the $q$--exponentials become usual exponentials.

\bigskip\bigskip

\noindent {\bf 4. The case of Euclidean and Galilei quantum groups.}

\medskip

In this last section we shall determine the $T$ matrices for two
inhomogeneous quantum groups, the Euclidean quantum group $E_q(2)$ and
the Galilei $\Gamma_q(1)$.

The quantizations of $E(2)$ have been throughly discussed in [9].
Here we shall be concerned with that quantum deformation, initially found
in [6], which can very simply be obtained by a contraction of the $SU_q(2)$
algebra, rescaling the generators as
$$P_+\,=\,\ee\, J_+\ , \quad~~\quad P_-\,=\,\ee\, J_-\ , \quad~~\quad
J\,=\,J_3\ ,$$
and taking the limit $\ee\rightarrow 0$.

If we define the new basis $(b_-,J,b_+)$, with
$$b_-\,=\,\esp{-zJ/2}\,P_-\,=\,\ee F\  , \quad\quad
b_+\,=\,\esp{zJ/2}\,P_+\,=\,\ee E$$
and denote by $\ (\pi_-,\pi,\pi_+)\ $ the dual basis of
$\ (b_-,J,b_+)\ $, the relations
$$\eqalign{
\dual{\pi_-}{b_-}\,=&\,1\,=\,\dual{\pi_-}{\ee\,F}=\dual{\ee\pi_-}{F}\ ,\cr
\dual{\pi}{J}\,=&\,1\,=\,\dual{\pi}{J_3}\ ,\cr
\dual{\pi_+}{b_+}\,=&\,1\,=\,\dual{\pi_+}{\ee\,E}=\dual{\ee\pi_+}{E}\ .\cr}$$
imply that $\pi_-=\ee^{-1}\,\phi,\ \
\pi=\gamma,\ \ \pi_+=\ee^{-1}\,\eta\ $, where $(\phi,\gamma,\eta)$ have
been introduced in section 2. By means of these rescalings
we can directly contract the relations of $\funq{SU(2)}$ obtaining:
$$[\pi_-,\pi_+]=0\ , \quad\quad [\pi,\pi_-]=-z\,\pi_-\ ,\quad\quad
[\pi,\pi_+]=-z\, \pi_+\ .\eqno{(16)}$$
Let us notice that the algebra relations in (10), (15) and (16) have
all the same structure.

In the limit $\ee\rightarrow 0$ the coproducts read
$$\eqalign{
m^*(\pi_-)=&\;\pi_-\otimes \uno + \esp{-\pi}\otimes \pi_-\ ,\cr
m^*(\pi)=&\;\uno\otimes \pi + \pi\otimes \uno\ ,\cr
m^*(\pi_+)=&\;\uno\otimes \pi_+ + \pi_+\otimes \esp{-\pi}\ ,\cr
}$$
and the antipodes
$$\eqalign{S^*(\pi_-)=&-\esp{\pi}\pi_-\ ,\cr
S^*(\pi)=&-\pi\ ,\cr
S^*(\pi_+)=&-\pi_+ \esp{\pi}\ . \cr
}$$
It is not difficult to show that this structure is equivalent to the one
found in [4,9,11]. It allows, however a very plain determination of the
universal $T$ matrix. Indeed, contracting the expression (13) we find
$$T=e^{\;\pi_-\otimes b_-}_z\ \ e^{\;\pi\otimes J}\ \ e^{\;\pi_+\otimes
b_+}_{-z}\ .\eqno{(17)}$$
Using the fact that
$$e_z^A\,e_z^B=e_z^{A+B}~~~~~~~{\rm if}~~~~~~~AB=e^{-z}BA\,,$$
all the properties of $T$ can be directly verified.

Let us finally analyze the results for the deformation of one dimensional
Galilei group, $\Gamma_q(1)$, which, as shown in [7], has remarkable physical
applications since it describes the dynamical symmetries of  magnon systems
on a linear lattice.
Contrary to what occurs for $E_q(2)$, the quantum group $\Gamma_q(1)$
cannot be obtained from a contraction: therefore, in order
to find the expression for the $T$ matrix we
have to determine the explicit duality relations.

The Hopf algebra of $\Gamma_q(1)$, [7], is defined by the commutation
relations
$$ \eqalign{
[B,P]\;=\;i M \,,\;\; ~~~&~~~ \;\;\; [B,T]\;=\;i/a \; \sin (a P)\,,\cr
            [P,T]\;=0 \,,\;\; ~~~&~~~ \;\;\; [M,\cdot] = 0\,, \cr}
$$
with coproducts, antipodes and counits
$$\eqalign{
  \Delta B = e^{-i\,aP} \otimes B + B \otimes e^{i \, aP} \; \; , & \;\;
  \Delta M = e^{-i\,aP} \otimes M + B \otimes e^{i \, aP} \cr
  \Delta P = \uno \otimes P + P \otimes \uno     \;\; ,      &   \; \;
  \Delta T = \uno \otimes T + T \otimes \uno \cr}
$$
$$
S(T) = -T \; ,         \;\;\;  S(B) = -B - a M \;, \;\;\;
S(P) = -P \; , \;\;\; S(M) = -M   \;\;.
$$

If we define the pairing
$$
\dual{\mu}{m} \; = \; \dual{x}{P} \; = \; \dual{t}{T} \; = \; \dual{v}{b} =
1 \; ,
$$
where $m = e^{-i a P}  M$ and $b= e^{i a P} B$,
we get the relations for $\funq{\Gamma(1)}$

$$ \eqalign{ [v,x]\;=\;2 i a v  \;\;\;\;\ ~~~&~~~ \;\;\; [v,\mu]\;=\;- a v^2
\cr
[x,\mu]\;=-2 i a \mu \;\;\; ~~~&~~~ \;\;\;  [t,.] = 0 \; , \cr}
$$
while the coproducts, antipodes and counits are as in the Lie case, namely
$$\eqalign{
  \Delta v = \uno \otimes v + v \otimes \uno \; \; , ~~&~~ \;\;
  \Delta \mu = \uno \otimes \mu + \mu \otimes \uno + i \, v \otimes x -
               1/2 \, v^2 \otimes t \cr
  \Delta t = \uno \otimes t + t \otimes \uno     \;\; ,      ~~&~~   \; \;
  \Delta x = \uno \otimes x + x \otimes \uno + i \, v \otimes t \cr}
$$
Defining the basis of $\Gamma_q(1)$ as
$X_{abcd} = m^{a} P^{b} T^{c}
b^{d}$ the dual basis of $\funq{\Gamma(1)}$ is
$\ (\mu^{a} x^{b} t^{c} v^{d})/(a\,!\,b\,!\,c\,!\,d\,!)$,
so that, like for the Heisenberg group, the universal $T$ matrix is
given in terms of simple exponential series
$$
 T \; = \; e^{\mu \otimes m} e^{x \otimes P} e^{t \otimes T}
e^{v \otimes b} \; .$$

To conclude we can mention that universal $T$ matrices for other
inhomogeneous quantum groups, as the singular deformation of $E(2)$, [9],
or the three dimensional Euclidean group, [5], can be obtained along the
same lines.
Obviously, in the presence of an $R$-matrix, the duality relations are
more easily determined. This is, for instance, the case of $\funq{E(3)}$,
whose relations
\footnote{${}^{(*)}$}{
We take here the opportunity of correcting a misprint of [5].},
in the notations of [5],
$$ \eqalign{
{}&[z,\tilde x] = - w\tilde x\,,\phantom{n(\theta/2)}\;~~~~~~~
[\theta,\tilde x] = w\,\sin\theta\,\tan(\theta/2)\,,\cr
{}&[\tilde y,\tilde x] = w \tilde y\,\tan(\theta/2)\,,~~~~~~~
[\theta,z] = w \,\sin\theta\,,\cr
{}&[\tilde y,z] = w\tilde y\,,\phantom{\tan(\theta/2)}~~~~~~~~
[\omega,\tilde y] = -2w\,\tan(\theta/2)\,,\cr}
$$
are defined in terms of (7).

\bigskip\bigskip

\noindent {\bf 5. Concluding remarks.}

\medskip

Starting from the definition of the universal $T$ matrix given in [1],
we have stressed its direct connection with the exponential
mapping for Lie groups. The  $T$ matrix collects both the structure
of the quantum algebra and of its dual and allows an algebraic
formulation of the standard ``$RTT$'' relations.
Its expression in terms of $q$--exponentials, as already given in [1],
is found to reduce to usual exponentials for those inhomogeneous
groups for which the quantum parameter can be reabsorbed by a new
definition of the generators of the quantum algebra
[5,9]: when this occurs, as for $E_\ell(2)$ and $\Gamma_q(1)$,
the parameter acquires a physical dimension and is naturally interpreted
as a lattice spacing. Therefore the universal $T$ matrix,
regarded in [1] as the quantum transfer matrix in
models of $(1+1)$ lattice field theory,
can also naturally be used for
the study of the quantum deformation of group properties  of
physical systems with kinematical symmetries generated by quantum
algebras, as in the cases of magnons and phonons [7,12].
In this context the explicit structure for all the semisimple quantum
groups and for
the other ones relevant for applications (e.g. $q$--Poincar\'e)
deserves a careful examination. Moreover
the use of those results in the study of non commutative geometry and
$q$--special function should also be relevant.

\bigskip\bigskip

\centerline{{\bf References.}}

\bigskip
\ii 1 C. Fronsdal and A. Galindo, Lett. Math. Phys. {\bf 27}, 59 (1993).
\smallskip
\ii 2 N.Yu. Reshetikhin, L.A. Takhtadzhyan and L.D. Faddeev, Leningrad
Math. J. {\bf 1}, 193 (1990).
\smallskip
\ii 3 S.L. Woronowicz, Comm. Math. Phys. {\bf 111}, 613 (1987) and
      Invent. Math {\bf 93}, 35 (1988).
\smallskip
\ii 4 S.L. Woronowicz, Lett. Math. Phys. {\bf 23},251 (1991), and
      Comm. Math. Phys. {\bf 144}, 417 (1992).
\smallskip
\ii 5 E. Celeghini, R. Giachetti, E. Sorace and M. Tarlini, J. Math. Phys.
      {\bf 31}, 2548 (1990), {\bf 32}, 1155, 1159 (1991), and
      ``{\it Contractions of quantum groups}'',
      in {\it Quantum Groups}, P.P. Kulish ed., Lecture Notes in Mathematics
      n. 1510, 221, (Springer-Verlag, 1992).
\smallskip
\ii 6 L.L. Vaksman and L.I. Korogodski, Sov. Math. Dokl. {\bf 39}, 173 (1989).
\smallskip
\ii 7 F. Bonechi, E. Celeghini, R. Giachetti, E. Sorace and M.
Tarlini, Phys. Rev. B {\bf 46}, 5727 (1992), J.Phys. A, {\bf 25},
L939 (1992).
\smallskip
\ii 8 M. Postnikov, {\it Lectures in Geometry, V}, Mir Publishers
(Moscow, 1986).
\ii 9 A. Ballesteros, E. Celeghini, R. Giachetti, E. Sorace and M.
Tarlini, {\it ``An R--matrix appoach to the quantization of the
euclidean group E(2)''}, J. Phys. A in press.
\smallskip
\jj {10} V.G. Drinfeld,
      ``{\it On some unsolved problems in Quantum Group theory}''
      in {\it Quantum Groups}, P.P. Kulish ed., Lecture Notes in Mathematics
      n. 1510, 221, (Springer-Verlag, 1992).
\smallskip
\jj {11} P. Schupp, P. Watts and B. Zumino, Lett. Math. Phys. {\bf 24},
141 (1992).
\smallskip
\ii {12} F. Bonechi, E. Celeghini, R. Giachetti, E. Sorace and M.
Tarlini, Phys. Rev. Lett. {\bf 68}, 3718 (1992).

\bye